\documentclass[twocolumn,showpacs,amsfonts,aps,prd,floatfix,nofootinbib,float,superscriptaddress]{revtex4} 
\usepackage{amsmath}
\usepackage{bm}
\usepackage{graphicx}
\usepackage{color}
\usepackage{booktabs}

\usepackage{placeins}
\usepackage{enumerate}

\begin{document}
\title{Stochastic potential and quantum decoherence of heavy quarkonium in the quark-gluon plasma}
\date{\today}

\author{Yukinao Akamatsu}
\affiliation{Kobayashi-Maskawa Institute for the Origin of Particles and the Universe (KMI), Nagoya University, Nagoya 464-8602, Japan}

\author{Alexander Rothkopf}
\affiliation{Fakult$\ddot a$t f$\ddot u$r Physik, Universit$\ddot a$t Bielefeld, D-33615 Bielefeld, Germany}

\begin{abstract}
We propose an open quantum systems approach to the physics of heavy quarkonia in a thermal medium, based on stochastic quantum evolution.
This description emphasizes the importance of collisions with the environment and focuses on the concept of spatial decoherence of the heavy quarkonium wave function.
It is shown how to determine the parameters of the dynamical evolution, i.e. the real potential and the noise strength, from a comparison with quantities to be obtained from lattice QCD.
Furthermore the imaginary part of the lattice QCD heavy quark potential is found to be naturally related to the strength of the noise correlations. We discuss the time evolution of $Q\bar{Q}$ analytically in a limiting scenario for the spatial decoherence and provide a qualitative 1-dimensional numerical simulation of the real-time dynamics.
\end{abstract}


\maketitle

\section{Introduction}\label{sec1}
The fate of heavy quarkonium states ($Q\bar Q$) at very high temperature is a long standing puzzle ever since suppression of $J/\Psi$ (the $c\bar c$ ground state with $J^P=1^-$) has been proposed as a prime signal for the formation of the deconfined state of QCD matter, the quark-gluon plasma (QGP) \cite{Matsui:1986dk,Hashimoto:1986nn}.
The production of $J/\Psi$ and $\Upsilon$ ($b\bar b$ states with $J^P=1^-$) in heavy-ion collisions has been measured in detail at the Relativistic Heavy Ion Collider (RHIC) \cite{Adare:2006ns} and more recently at the Large Hadron Collider (LHC) \cite{MartinezGarcia:2011nf,Chatrchyan:2011pe}. 
Partial suppression of $J/\Psi$ \cite{Adare:2006ns} and the relative suppression of excited states such as $\Upsilon(\rm 2S,3S)$ in comparison to $\Upsilon(\rm 1S)$ \cite{Chatrchyan:2011pe} have indeed been observed.

In order to attack the question of $Q\bar{Q}$ survival, several approaches have been deployed. 
Maximum Entropy (MEM) analyses \cite{Asakawa:2000tr} of $J/\Psi$ and $\Upsilon$ spectral functions in lattice QCD have revealed a possible survival of these $c\bar{c}$ and $b\bar{b}$ bound states up to around $T\sim 2T_c$ with $T_c$ being the critical temperature for the hadron-quark transition \cite{Asakawa:2003re}.
The observed spectral structures are interpreted as a sign of $J/\Psi$ survival by some authors \cite{Asakawa:2003re}, while for others they represent mere threshold enhancement \cite{Mocsy:2007yj}. 
The unclear definition of what constitutes a heavy quark bound state is clearly a drawback of this otherwise solid approach.

On the other hand, purely real potential models, based on quantities, such as the difference in the free or internal energies \cite{Kaczmarek:2005jy} of a medium with and without a $Q\bar{Q}$ inserted, are struggling with the absence of a Schr\"odinger equation derived from QCD.
This leads to ambiguities in their definition of a heavy quark potential. 
Progress has been made in deriving a Schr\"odinger equation from QCD by evaluating the late time evolution of the real-time thermal Wilson loop \cite{Laine:2006ns,Beraudo:2007ky,Brambilla:2008cx}.
An initial perturbative study at very high temperatures, based on the Hard-Thermal-Loop approximation, yielded a complex potential, whose imaginary part is induced by collisions with the light plasma particles.
Recently \cite{Rothkopf:2011db} showed how to relate the potential to the spectral decomposition of the thermal Wilson loop and based on lattice QCD simulations confirmed the existence of an imaginary part above $T_c$.

In this paper, we instead propose an open quantum systems approach (see \cite{Young:2010jq,Borghini:2011yq} for earlier attempts), using a stochastic, i.e. a fully dynamical description of $Q\bar Q$ in the QGP.
The noise inherent in our stochastic treatment is a result of interactions between the $Q\bar{Q}$ open system and the medium, which is traced out from the description.
It not only naturally gives a physical meaning to the imaginary part found in the heavy $Q\bar{Q}$ potential from lattice studies but also provides an indispensable ingredient in discussing the time evolution of $Q\bar Q$, i.e. spatial decoherence of the wave function due to the collisions with plasma particles. 

Without spatial decoherence, such as in real potential models, the effects of collisions with the environment are underestimated, whereas having an explicit imaginary part in the potential will lead to an unabated suppression of all available states.
Both of these situation are expected to be unphysical and we will show how our proposed model fills this gap in the following sections.

Our investigation makes close contact with recent lattice QCD studies \cite{Rothkopf:2011db} by relating the strength of the noise to the spectral function of the thermal Wilson loop, thus making possible the determination of the required parameters at all temperatures non-perturbatively. 

This paper is organized as follows.
In section \ref{sec2}, we formulate the stochastic equation of motion for the heavy quarkonium wave function and give a physical interpretation of the correlation present in the noise term. The limitations of our proposed formulation are listed and discussed.
In order to connect the stochastic model to actual physics, we show in Sec.~\ref{sec3} how to determine some of the basic parameters of the stochastic dynamics from first principles lattice QCD simulations.
In Sec.~\ref{sec4}, we derive a master equation for the density matrix of heavy quarkonium states from the underlying stochastic time evolution.
We then trace out the global motion from the density matrix, which enables us to reduce the two-body problem to a one-body problem.
In Sec.~\ref{sec5}, we first list several criteria for bound state survival and determine their values in one dimensional numerical simulations of the stochastic dynamics based on different parameter sets adapted from the literature.
Finally in Sec.~\ref{sec6}, we summarize our work and give an outlook toward how to arrive at a more complete understanding of the heavy quarkonium states in the QGP.

\section{Stochastic Evolution of an Open Quantum System}\label{sec2}
Suppose we insert a heavy quark bound state into the thermal medium and let it evolve in time.
In a situation where a classical description is adequate, the relevant dynamics of the heavy particle is governed by Langevin's theory of Brownian motion.
On the other hand, in a situation where quantum effects play a crucial role, the dynamics should be described by the theory of open quantum systems \cite{open_quantum_system}.
In the latter case, interactions with the light particles of the heat bath lead to fluctuations in the time evolution of the heavy particle wave function, or in other words, to fluctuations of the Hamiltonian, i.e. the potential.
In the phenomenologically relevant region of the strongly coupled quark-gluon plasma just above the deconfinement phase transition, surely the latter approach is required.

The full quantum system is described by a density matrix and its time evolution is governed by the full Hamiltonian (in the Schr\"odinger picture):
\begin{eqnarray}
&& H = H_{\rm sys} \otimes I_{\rm med} + I_{\rm sys}\otimes H_{\rm med}+H_{\rm int},\\
&& \frac{d}{dt} \rho(t)=\frac{1}{i\hbar}\left[H,\rho(t)\right], 
\end{eqnarray}
with some initial condition 
$\rho(0)=\rho_{\rm sys}\otimes\rho_{\rm med}$.
In the following, we are only interested in the physics of the subsystem, which is described by a reduced density matrix $\rho_{\rm sys}(t)\equiv {\rm Tr}_{\rm med}\{\rho(t)\}$.
The dynamics of $\rho_{\rm sys}(t)$ is often unraveled into a stochastic evolution in Hilbert space. It can be described e.g. by a master equation of Lindblad-form \cite{Lindblad:1975ef}, corresponding to stochastic quantum evolution that incorporates quantum state diffusion \cite{Gisin:1992} or quantum jump processes \cite{Plenio:1997ep}.

In our approach we first wish to formulate a wave function based description of the stochastic time evolution of a non-relativistic heavy quarkonium state, whose pair annihilation is neglected. We use the short hand notation $\bm{X} \equiv (\bm x_1,\bm x_2)$ to denote the position of the constituent quarks, written as single vector in a 6-dimensional space.
The corresponding distance in this vector space reads $\Delta X\equiv |\bm X -\bm X'|=\sqrt{(\bm x_1-\bm x'_1)^2+(\bm x_2-\bm x'_2)^2}$.
The stochastic time evolution operator $U^{(X)}_{\Theta}(t|0)$ has to be unitary and we assume that the dynamics it describes is both Markovian and linear in the heavy quarkonium wave function $\Psi_{Q\bar{Q}}(\bm X,t)$.
Time evolution is thus described by
\begin{eqnarray}
\label{eq:evolution}
\Psi_{Q\bar{Q}}(\bm X,t)&=&U^{(X)}_{\Theta}(t|0)\Psi_{Q\bar{Q}}(\bm X,0),\\
U^{(X)}_{\Theta}(t|0)&=&{\rm T}\exp\left[
-\frac{i}{\hbar}\int^t_0 dt'
\left\{H(\bm X)+\Theta(\bm X,t')
\right\}
\right],\nonumber \\
H(\bm X)&\equiv&2M-\frac{\hbar^2\nabla_X^2}{2M}+V(\bm X),\nonumber
\end{eqnarray}
where $\rm T$ denotes the time-ordered product and the stochastic term in the Hamiltonian $\Theta(\bm X,t)$ corresponds to Gaussian white noise with the characteristics:
\begin{eqnarray}
\label{eq:fluctuation}
\langle \Theta(\bm X,t)\rangle=0, \ \langle \Theta(\bm X,t)\Theta(\bm X',t')\rangle=\hbar\Gamma(\bm X,\bm X')\delta_{tt'}/\Delta t.
\end{eqnarray}
Here we choose as a time scale $\Delta t$, during which the plasma particles experience many collisions but the heavy quarkonium state does not change considerably.
Such intermediate time scale $\Delta t$ is expected to exist due to the hierarchy in the mass scale $M\gg T$.
The fluctuation of the Hamiltonian derives from the variations in the transferred energy during $\Delta t$ originating from collisions with the medium particles.
The squared strength of the fluctuations at $\bm X$, which is given by $\hbar\Gamma(\bm X,\bm X)/\Delta t$, represents how frequently and effectively the energy transfer takes place between the heavy quarkonium and the surrounding medium.
In a uniform and isotropic system, which we consider, $\Gamma(\bm X,\bm X)$ depends only on $|\bm x_1-\bm x_2|$.
Since the relativistic medium particles are of size $l_{\rm th} \equiv 2\pi\hbar c/k_B T$, the fluctuations, or equivalently the transferred energy, is expected to be correlated at $\bm X$ and $\bm X'$ if $\Delta X<l_{\rm th}$.
Therefore we expect that $\Gamma(\bm X,\bm X')\sim 0$ if $\Delta X$ is larger than $\sim l_{\rm th}$.

We take as cutoff length scale for this description the Compton wavelength of the heavy quarks $l_{\rm cut}=\hbar/Mc\ll l_{\rm th}$, so that the non-relativistic treatment is valid.
Since the typical momentum of heavy quarks near thermal equilibrium is $\sqrt{k_BMT}\equiv 2\pi\hbar/l_{Q,\rm th}$, there is a hierarchy $l_{\rm cut}\ll l_{Q,\rm th}\ll l_{\rm th}$.
It is this hierarchy that requires the effective stochastic dynamics of heavy quarks to be unitary, because collision processes have almost no chance to excite heavy quarks to a very high momentum state beyond the cutoff momentum scale $2\pi\hbar/l_{\rm cut}$.Note
that this argument implies that a mechanism
for thermalization must also be incorporated into a
complete effective dynamics of heavy quarks.

Given the manifestly unitary evolution in Eq.\eqref{eq:evolution}, the stochastic differential equation for the wave function can be written down by expanding $U^{(X)}_{\Theta}(t+\Delta t|t)$ in $\Delta t$, yielding
up to $\mathcal O(\Delta t^{3/2})$
\begin{eqnarray}
\label{eq:expansion}
&&U^{(X)}_{\Theta}(t+\Delta t|t) \\
&& \ \ \approx1-\frac{i\Delta t}{\hbar}\left\{H(\bm X)+\Theta(\bm X,t)\right\}
-\frac{(\Delta t)^2}{2\hbar^2}\Theta(\bm X,t)^2 \nonumber \\
&& \ \ =1-\frac{i\Delta t}{\hbar}\left\{H(\bm X)-\frac{i}{2}\Gamma(\bm X,\bm X)\right\}
-\frac{i\Delta t}{\hbar}\Xi(\bm X,t),\nonumber \\
&&\Xi(\bm X,t)\equiv\Theta(\bm X,t)-\frac{i\Delta t}{2\hbar}\left\{\Theta(\bm X,t)^2
-\langle \Theta(\bm X,t)^2\rangle\right\}. \nonumber
\end{eqnarray}
Here we define a new complex noise $\Xi(\bm X,t)$ in terms of $\Theta(\bm X,t)$ so that $\langle \Xi(\bm X,t)\rangle=0$.
The stochastic differential equation for the wave function in Ito discretization thus reads
\begin{eqnarray}
\label{eq:schroedinger}
&&i\hbar\frac{\partial}{\partial t}\Psi_{Q\bar{Q}}(\bm X,t)\\
&& \ \ \ =\Bigl\{H(\bm X)
-\frac{i}{2}\Gamma(\bm X,\bm X)+\Xi(\bm X,t)\Bigr\}\Psi_{Q\bar{Q}}(\bm X,t).\nonumber
\end{eqnarray}
Note that the expansion of the unitary time evolution operator has induced what appears to be a non-unitarity in the Schr\"odinger equation.
Since the origin of this contribution however are the noise terms of Eq.\eqref{eq:expansion}, it is only after taking the ensemble average that damping of the wave function is observed.

We would like to remark that the formulation of the stochastic dynamics based on Eq.\eqref{eq:evolution} is most likely not able to describe the quantum analog to the drag force in classical Langevin theory. Classically, this corresponds to an absence of a friction term for momentum diffusion, i.e. the momentum distribution of the heavy quarks diffuses without resistance. The effects related to these absent terms become important to the dynamics once the typical energy
of the heavy quarks reaches and exceeds the temperature.
Hence, roughly only for time scales smaller than
the heavy quark relaxation time our description
should be valid. Note that heavy quark relaxation
time can be much longer than that of the medium $\Delta t$ by a factor $\sim M/T$, so that the applicability of our description is not as severely limited as it might appear at first sight.

The above derivation of Eq.\eqref{eq:schroedinger} is based on the expansion of $U^{(X)}_{\Theta}(t+\Delta t|t)$ in $\Delta t$.
However, since the time step is related to physical scales as $\Delta t\sim\hbar/g_s^4k_BT$, it may not be an expansion in small {\it dimensionless} number.
This subtlety is overcome by introducing $N$ independent Gaussian noise variables $\Theta_i(\bm X,t)$ $(i=1,\cdots,N)$ which satisfies Eq.\eqref{eq:fluctuation} with $\Delta t$ replaced by $\Delta t/N$.
If we define $\Theta(\bm X,t)\equiv\frac{1}{N}\sum_{i=1}^{N}\Theta_i(\bm X,t)$, $\Theta(\bm X,t)$ satisfies the original definition Eq.\eqref{eq:fluctuation}.
The new noise $\Theta_i(\bm X,t)$ has a natural interpretation as fluctuation of the Hamiltonian in Eq.\eqref{eq:evolution} with artificial discretization time scale $\Delta t/N$.
We then assume that Eq.\eqref{eq:evolution} can still be expanded as time-ordered product of white noise, now with respect to the $\Theta_i(\bm X,t)$'s. Taking the noise to be independent even on time scales smaller than $\Delta t$ is a prerequisite to obtaining a Markovian master equation of the form present in Section IV.
As we observe heavy quarkonium on time scales longer than $\Delta t$, we discretize Eq.\eqref{eq:evolution}, expand Eq.\eqref{eq:expansion}, and simulate Eq.\eqref{eq:schroedinger} on a smaller timescale $\Delta t/N$ instead of $\Delta t$.
In the remaining part of this paper, we distinguish $dt\equiv\Delta t/N$ from $\Delta t$.

\section{Spectral Analysis}\label{sec3}

Now that we have formulated the stochastic evolution, one has to ask how the unknown functions $V(\bm X)$ and $\Gamma(\bm X,\bm X')$ in Eq.\eqref{eq:schroedinger} can be determined. Since we are interested in describing physics in the strongly coupled region of the quark-gluon plasma, we are urged to find non-perturbative means to do so.

In analogy to the concept of the Nambu-Bethe-Salpeter (NBS) wave function at zero temperature, the averaged wave function $\Psi_{Q\bar{Q}}(\bm X,t)$ of heavy quarkonium at finite temperature is identified by a suitable matching prescription with the following gauge invariant mesonic correlator ($t>0$):
\begin{eqnarray}
\label{eq:correlator}
D^>(\bm X,t)=\langle M(\bm X,t)M^{\dagger}(\bm X_0,0)\rangle_T
\equiv\langle \Psi_{Q\bar{Q}}(\bm X,t)\rangle_{_{\Xi}},
\end{eqnarray}
where $M^{(\dagger)}(\bm X,t)$ is the point-split heavy quarkonium annihilation (creation) operator and $\langle \cdot \rangle_T$ denotes the field theoretical average over the medium degrees of freedom at finite temperature\cite{Laine:2006ns,Rothkopf:2011db}.
In the case of infinitely heavy quark masses, where $m_Q\gg T$ and $m_Q\gg\Lambda_{\rm QCD}$, the left hand side of Eq.\eqref{eq:correlator} can be calculated in terms of the thermal Wilson loop 
\begin{align}
 D^>_{m_Q\to\infty}(\bm X,t) = \langle W(\bm X,t) \rangle_T = \langle {\rm exp}\Big[-\frac{i}{\hbar} \oint dy^\mu A_\mu(y)\Big]\rangle_T
\end{align}
where $A^\mu(x)$ denotes the gauge field of the medium gluons, which is integrated along the rectangular contour connecting $\bm X_0$ at $t=0$ and $\bm X$ at $t$.
The right hand side of Eq.\eqref{eq:correlator} becomes
\begin{eqnarray}
\langle \Psi_{Q\bar{Q}}(\bm X,t)\rangle_{_{\Xi}}\propto\exp\left[
-\frac{it}{\hbar}\left\{
V(\bm X)-\frac{i}{2}\Gamma(\bm X,\bm X)
\right\}
\right].\label{eq:dampingWFavg}
\end{eqnarray}
In order to continue we have to remember that all equations appearing so far are expressed in real-time $t$. 
As we wish to use Monte-Carlo simulations of lattice QCD, formulated in Euclidean time to evaluate the Wilson loop non-perturbatively
\cite{footnote1}, their results need to be analytically continued.
One possible strategy is to use spectral functions, defined from the forward and backward propagator
\begin{align}
\sigma(\bm X,\omega)= D^>(\bm X,\omega) - D^<(\bm X,\omega)
\end{align}
Since the backward correlator for infinitely heavy quarks vanishes $D^<_{m_Q\to\infty}(\bm X,t)=0$, one finds that the spectral function $\sigma(\bm X,\omega)$ is nothing but the Fourier transform of the forward correlator $D^>(\bm X,t)$ and thus of the real-time Wilson Loop. Analytic continuation yields
\begin{align}
\label{eq:continuation}
W(\bm X,\tau)=\int_{-\infty}^\infty d\omega\;e^{-\omega \tau}  \sigma(\bm X,\omega),
\end{align}
which tells us that the relevant spectral information can be extracted from imaginary time data accessible on the lattice. 
In practice the numerical determination of the above spectral function from the Euclidean Wilson loop is based on a form of Bayesian inference called Maximum Entropy Method (MEM) \cite{Asakawa:2000tr}. 

In case that the spectrum of the Wilson loop exhibits well defined structures, e.g. peaks of Breit-Wigner form \cite{footnote2}, the real potential $V(\bm X)$ and the local noise correlations $\Gamma(\bm X,\bm X)$ of Eq.\eqref{eq:dampingWFavg} are readily obtained from the position and width of the lowest lying peak \cite{Rothkopf:2011db}:
\begin{eqnarray}
\label{eq:Breit-Wigner}
\sigma(\bm X,\omega)\propto
\frac{\Gamma(\bm X,\bm X)/2}
{\left(\hbar\omega-V(\bm X)\right)^2
+\left(\Gamma(\bm X,\bm X)/2\right)^2}.
\end{eqnarray}

In the context of our model, the width of the spectral function is interpreted as uncertainty in the actual value of the real potential in the open system.
Consequently time evolution is always unitary for each realization of the stochastic ensemble.
In this sense it characterizes the decoherence between the initial and current wave function.
This is in contrast to the deterministic notion of the width as an imaginary part, i.e. an explicit contribution to the Hamiltonian, which violates hermiticity and leaves open questions about conservation of the probability density.

Even though we have already obtained two main ingredients of our model, the real potential and the local noise, we are still missing another piece of the puzzle, i.e. information on the off-diagonal noise terms.
A quantity that might allow extracting this set of parameters is introduced in the next section.
As a first ansatz, to allow simulations based on Eq.\eqref{eq:schroedinger}, one might assume that such spatial correlations within the fluctuations, i.e. the $\Delta X$-dependence of $\Gamma(\bm X,\bm X')$, will decay with the thermal wavelength as $\exp(-(\Delta X)^2/l_{\rm th}^2)$.

\section{The Master Equation}\label{sec4}
In principle, one can simulate the stochastic time evolution in Eq.\eqref{eq:schroedinger} directly, but in order to precisely define the time evolution of the heavy quark bound state in a thermal medium, we are urged to go one step further in our analysis.
To this end we derive a master equation for the reduced density matrix of states $\rho_{Q\bar{Q}}(\bm X,\bm X',t) \equiv\langle \Psi_{Q\bar{Q}}(\bm X,t)\Psi^*_{Q\bar{Q}}(\bm X',t)\rangle_{_{\Xi}}$.
Noting that an infinitesimal time step in the time evolution of $\rho_{Q\bar{Q}}(\bm X,\bm X',t)$ is given by
\begin{eqnarray}			
&&\rho_{Q\bar{Q}}(\bm X,\bm X',t+dt) \\
&& \ \ \ =\langle U^{(X)}_{\Theta}(t+dt|t)U^{(X')*}_{\Theta}(t+dt|t)\rangle_{_{\Xi}} \ \rho_{Q\bar{Q}}(\bm X,\bm X',t),\nonumber
\end{eqnarray}
and calculating $\langle U^{(X)}_{\Theta}(t+dt|t) U^{(X')*}_{\Theta}(t+dt|t)\rangle_{_{\Xi}}$ using Eqs.\eqref{eq:fluctuation} and \eqref{eq:expansion}, it is straightforward to derive the following relation:
\begin{eqnarray}
\label{eq:master}
\frac{\partial \rho_{Q\bar{Q}}(\bm X,\bm X',t)}{\partial t}&=&
\frac{H(\bm X)-H(\bm X')}{i\hbar}\rho_{Q\bar{Q}}(\bm X,\bm X',t)\\
&&+\frac{F(\bm X,\bm X')}{\hbar}\rho_{Q\bar{Q}}(\bm X,\bm X',t),\nonumber \\
F(\bm X,\bm X')&\equiv&
\Gamma\left(\bm X,\bm X'\right)
-\frac{\Gamma(\bm X,\bm X)+\Gamma(\bm X',\bm X')}{2}.\nonumber
\end{eqnarray}
The conservation of the trace ${\rm Tr_{\it X}} \{\rho_{Q\bar{Q}} (\bm X,\bm X',t)\}\equiv \int d^6X \rho_{Q\bar{Q}}(\bm X,\bm X,t)=1$ can be proved easily.
Since the time evolution of this averaged quantity now depends also explicitly on the off-diagonal spatial correlations in the noise, it appears as a promising candidate to extract the yet undetermined values of $\Gamma(\bm X',\bm X')$.
Future work thus needs to focus on expressing the density matrix of states for the heavy quark system as field theoretical operator amenable to lattice QCD simulations.

An equivalent master equation is known from the studies of scattering models and applications of influence-functional techniques to random potentials \cite{Gallis:1990}.
Its applicability to the physics of heavy quarks, however, has so far not been exploited.
Note that in comparison, the recently proposed quantum optical master equation \cite{Borghini:2011yq} is applicable where the rotating wave approximation is valid, i.e. in discussing transitions between deeply bound states \cite{open_quantum_system}.
Our approach on the other hand is geared toward the description of loosely bound states and melting phenomena, since it explicitly describes both constituting particles individually.

As we are interested in the relative motion of the heavy quarks we trace out the global motion.
Defining the further reduced density matrix $\hat \rho_{Q\bar{Q}}(\bm r,\bm r',t)\equiv{\rm Tr}_{R}\{\rho_{Q\bar{Q}}(\bm X_{R,r},\bm X_{R',r'},t)\}=\int d^3R\rho_{Q\bar{Q}}(\bm X_{R,r},\bm X_{R,r'},t)$, where $\bm X_{R,r}\equiv(\bm R+\bm r/2,\bm R-\bm r/2)$, we obtain from Eq.\eqref{eq:master} the master equation for $\hat\rho_{Q\bar{Q}}(\bm r,\bm r',t)$:
\begin{eqnarray}
\label{eq:master2}
\frac{\partial \hat\rho_{Q\bar{Q}}(\bm r,\bm r',t)}{\partial t}&=&
\frac{h(\bm r)-h(\bm r')}{i\hbar}\hat\rho_{Q\bar{Q}}(\bm r,\bm r',t)\nonumber \\
&&+\frac{f(\bm r,\bm r')}{\hbar}\hat\rho_{Q\bar{Q}}(\bm r,\bm r',t),
\end{eqnarray}
where $h(\bm r)\equiv 2M-\frac{\hbar^2\nabla_r^2}{M}+v(r)$ and
$f(\bm r,\bm r')\equiv \gamma(\bm r,\bm r')-\{\gamma(\bm r,\bm r)+\gamma(\bm r',\bm r')\}/2$.
Here we define $v(r)\equiv V(\bm X_{0,r})$ and $\gamma(\bm r,\bm r')\equiv\Gamma(\bm X_{0,r},\bm X_{0,r'})$ using translational and rotational invariance.
Correspondingly, the stochastic dynamics of the relative coordinates of the heavy quark pair is given in Ito discretization by 
\begin{eqnarray}
\label{eq:schroedinger2}
&& i\hbar\frac{\partial}{\partial t}\psi_{Q\bar{Q}}(\bm r,t)\\
&& \ \ \ =\Bigl\{h(\bm r)
-\frac{i}{2}\gamma(\bm r,\bm r)+\xi(\bm r,t)\Bigr\}\psi_{Q\bar{Q}}(\bm r,t),\nonumber\\
&&\xi(\bm r,t)\equiv \theta(\bm r,t)-\frac{idt}{2\hbar}\left\{\theta(\bm r,t)^2
-\langle \theta(\bm r,t)^2\rangle\right\},\nonumber
\end{eqnarray}
with $\langle\theta(\bm r,t)\rangle=0$ and $\langle \theta(\bm r,t)\theta(\bm r',t')\rangle=
\hbar\gamma\left(\bm r,\bm r'\right)\delta_{tt'}/dt$.

In order to discuss the time evolution of heavy quarkonium in the next section, we can take two equivalent, while differently nuanced standpoints. From the theoretical side, one wishes to describe the time evolution of the system by as simple equations as possible, which is best done in the basis of eigenstates of the Hamiltonian governing the dynamics $h(\bm r)$. On the other hand to build a bridge to experiment and its notion of $Q\bar{Q}$ suppression, one asks how much of an initial bound state has survived after a certain time $t$. This question is best answered when the evolution of the system is observed in the basis of the eigenstates of the initial vacuum Hamiltonian $h^{\rm vac}(\bm r)$ that governs the bound state.

We thus transform Eq.\eqref{eq:master2} by expanding
\begin{align}
 \hat\rho_{Q\bar{Q}}(\bm r,\bm r',t)&=\sum_{nm}c^{\rm m}_{nm}(t)\psi_n(\bm r)\psi_m^*(\bm r')\\
&=\sum_{nm}c^{\rm v}_{nm}(t)\phi_n(\bm r)\phi_m^*(\bm r')
\end{align}
in a complete set of eigenfunctions $\psi_n(\bm r)$ (with eigenvalue $\epsilon_n$) of $h(\bm r)$ or $\phi_n(\bm r)$ (with eigenvalue $e_n$) of $h^{\rm vac}(\bm r)$.
The expansion coefficients are thus given by
\begin{align}
\nonumber c^{\rm  m}_{nm}=\int d^3r d^3r'\; \psi^*_n (\bm r)\langle\psi_{Q\bar{Q}}(\bm r,t)\psi_{Q\bar{Q}}^*(\bm r',t)\rangle_{\xi} \psi_m (\bm r')\\
c^{\rm v}_{nm}=\int d^3r d^3r'\; \phi^*_n (\bm r)\langle\psi_{Q\bar{Q}}(\bm r,t)\psi_{Q\bar{Q}}^*(\bm r',t)\rangle_{\xi} \phi_m (\bm r')%}
\end{align}
In terms of the in-medium state admixtures $c^{\rm m}_{nm}$, the time evolution of the system according to Eq.\eqref{eq:master} takes on the following form:
\begin{eqnarray}
\label{eq:rate}
\dot c^{\rm m}_{nm}(t)&=&\frac{\epsilon_n-\epsilon_m}{i\hbar}c^{\rm m}_{nm}(t)
+\frac{1}{\hbar}\sum_{kl}\gamma_{nk,lm}c^{\rm m}_{kl}(t)\nonumber\\
&&-\frac{1}{2\hbar}\sum_k\left\{\gamma_{nk}c^{\rm m}_{km}(t)+c^{\rm m}_{nk}(t)\gamma_{km}\right\},
\end{eqnarray}
where
$\gamma_{nk,lm}\equiv
\int d^3r d^3r'\psi^*_n (\bm r)
\psi^*_l (\bm r')\gamma\left(\bm r,\bm r'\right)
\psi_k(\bm r)\psi_m(\bm r')$
and 
$\gamma_{nm}\equiv\int d^3r\psi^*_n(\bm r)\gamma(\bm r,\bm r)\psi_m(\bm r)$.
Here $c^{\rm m}_{nm}(t)$ and $\gamma_{nm}$ are hermitian matrices and $\gamma_{nk,lm}=\gamma_{lm,nk}=\gamma_{kn,ml}^*$.

\section{Stochastic Evolution of Heavy Quarkonium in the QGP}\label{sec5}

To model the physics of heavy quarkonia related to relativistic heavy ion collisions at RHIC and the LHC, we use the following setup.
The initial heavy quarkonium is taken as the ground state of the {\it vacuum} Hamiltonian $h^{\rm vac}(r)$, characterized by a Coulombic and linearly rising potential at small and large separation distances respectively.
Once such a bound state enters the region in which a quark-gluon plasma is present, the interactions with this hot medium are taken into account through time evolution as given by Eq.\eqref{eq:schroedinger2} with appropriate model parameters $v(r)$ and $\gamma(r,r')$ corresponding to temperatures $T>T_c\sim$170MeV.
Depending on whether we wish to focus on the time evolution of the in-medium states or the initial bound-states in the QGP, we may use either the medium or vacuum bound state survival probabilities
\begin{align}
 P^{\rm m}(t)&\equiv\sum_{1\leq n\leq N^{\rm m}_{\rm b}} c^{\rm m}_{nn}(t)\\
 P^{\rm v}(t)&\equiv\sum_{1\leq n\leq N^{\rm v}_{\rm b}} c^{\rm v}_{nn}(t),
\end{align}
where $N^{\rm m}_{\rm b}$ ($N^{\rm v}_{\rm b}$)is the number of bound states of $h(x)$ ($h^{\rm vac}(x)$).

\subsection{Analytic Considerations}

Among three different representations of the stochastic dynamics in Eqs.\eqref{eq:master2},\eqref{eq:schroedinger2}, and \eqref{eq:rate}, the qualitative feature of $P^{\rm m}(t)$ can be discussed most clearly by using Eq.\eqref{eq:rate}. The following arguments hold also for $P^{\rm v}(t)$ as long as $h(\bm r)=h^{\rm vac}(\bm r)$.

In the simplest case where $\gamma(\bm r,\bm r')=\gamma$, we can show that $\gamma_{nk,lm}=\gamma\delta_{nk}\delta_{lm}$,
$\gamma_{nk}=\gamma\delta_{nk}$, and thus 
$\dot c^{\rm m}_{nm}(t)=\frac{\epsilon_n-\epsilon_m}{i\hbar}c^{\rm m}_{nm}(t)$,
$\dot P^{\rm m}(t)=0$. 
Therefore with spatially uniform fluctuations, quarkonium survival is completely determined by $h(x)$, which is the case for the purely real potential models.

On the other hand, in a case where the spatial correlations factor according to $\gamma(\bm r,\bm r')=\gamma \exp (-(\Delta r)^2/2l_{\rm corr}^2)$ with $\Delta r\equiv |\bm r-\bm r'|$ and $l_{\rm corr}\ll l_{\rm cut}$, we can show that $\gamma_{nk}=\gamma\delta_{nk}$ and $\gamma_{nk,lm}\rightarrow 0$ as $l_{\rm corr}/l_{\rm cut}\rightarrow 0$.
The latter relation is obtained by using the fact that the eigenfunction $\psi_n(x)$ varies over length larger than $l_{\rm cut}$:
\begin{eqnarray}
\gamma_{nk,lm}&\approx&
\gamma\int d^3r d^3r' \psi^*_n (\bm r)\psi^*_l (\bm r)
e^{-\frac{|\bm r-\bm r'|^2}{2l_{\rm corr}^2}}\psi_k(\bm r)\psi_m(\bm r)\nonumber \\
&<&\gamma(\sqrt{2\pi}l_{\rm corr}/l_{\rm cut})^3
\longrightarrow 0 \ \ \ (l_{\rm corr}/l_{\rm cut}\rightarrow 0).
\end{eqnarray}
Then we obtain 
$\dot c^{\rm m}_{nm}(t)=\frac{\epsilon_n-\epsilon_m}{i\hbar}c^{\rm m}_{nm}(t)-\frac{\gamma}{\hbar}c^{\rm m}_{nm}(t)$, $\dot P^{\rm m}(t)=-\frac{\gamma}{\hbar}P^{\rm m}(t)$.
Therefore with very localized fluctuation, the damping is determined solely by $\gamma=\gamma(\bm r,\bm r)$.
Since the correlation length $l_{\rm corr}$ of the fluctuation is below the cutoff scale $l_{\rm cut}$, such fluctuations excite wave modes which are beyond the model description, i.e. all the diagonal elements of $c^{\rm m}_{nm}(t)$ decay as $\sim e^{-\gamma t/\hbar}$ and 
${\rm Tr}_{r}\{\hat\rho(r,r',t)\}$ is not conserved.
In the physical case $l_{\rm corr}=l_{\rm th}$, the above arguments suggest that the states with typical scale larger than $l_{\rm th}$ decay while those smaller than $l_{\rm th}$ remain undisturbed, supporting the argument that the fate of bound states in essence is a hadronic thermometer. 

\subsection{1d Numerical Simulations}

For a first qualitative exploration of the above ideas, we carry out one-dimensional numerical simulations of the stochastic dynamics at $T\simeq 2.33T_c$, based on different parameter sets adapted from a lattice QCD study \cite{Rothkopf:2011db} (models (A) and (C)), the perturbative (PT) complex potential calculated in \cite{Laine:2006ns} (model (B)) and the classic Debye screening scenario of \cite{Matsui:1986dk} (model (D)). (Throughout this section, we adopt the natural unit $\hbar=c=k_B=1$.)

\begin{center}
 \begin{tabular}{cccc}
   Model            & $v(x)$    & $\gamma(x,x)$ & $l_{\rm corr}$  \\
\hline\hline
   (A)  & $v^{\rm vac}(x)$        & $\sigma |x|$             & $dx$  \\
   (B)  & \ Re$[v^{\rm PT}(x)]$ \ & \ Im$[v^{\rm PT}(x)] \ $ & \ $dx$ \ \\
   (C)  & $v^{\rm vac}(x)$        & $\sigma |x|$             & $4dx$ \\
   (D)  & $- \frac{\alpha e^{-m_D|x|}}{|x|}$        & -            &  - \\
 \end{tabular}
\end{center}

\begin{figure}[b!]
\includegraphics[scale=0.33,angle=-90,clip]{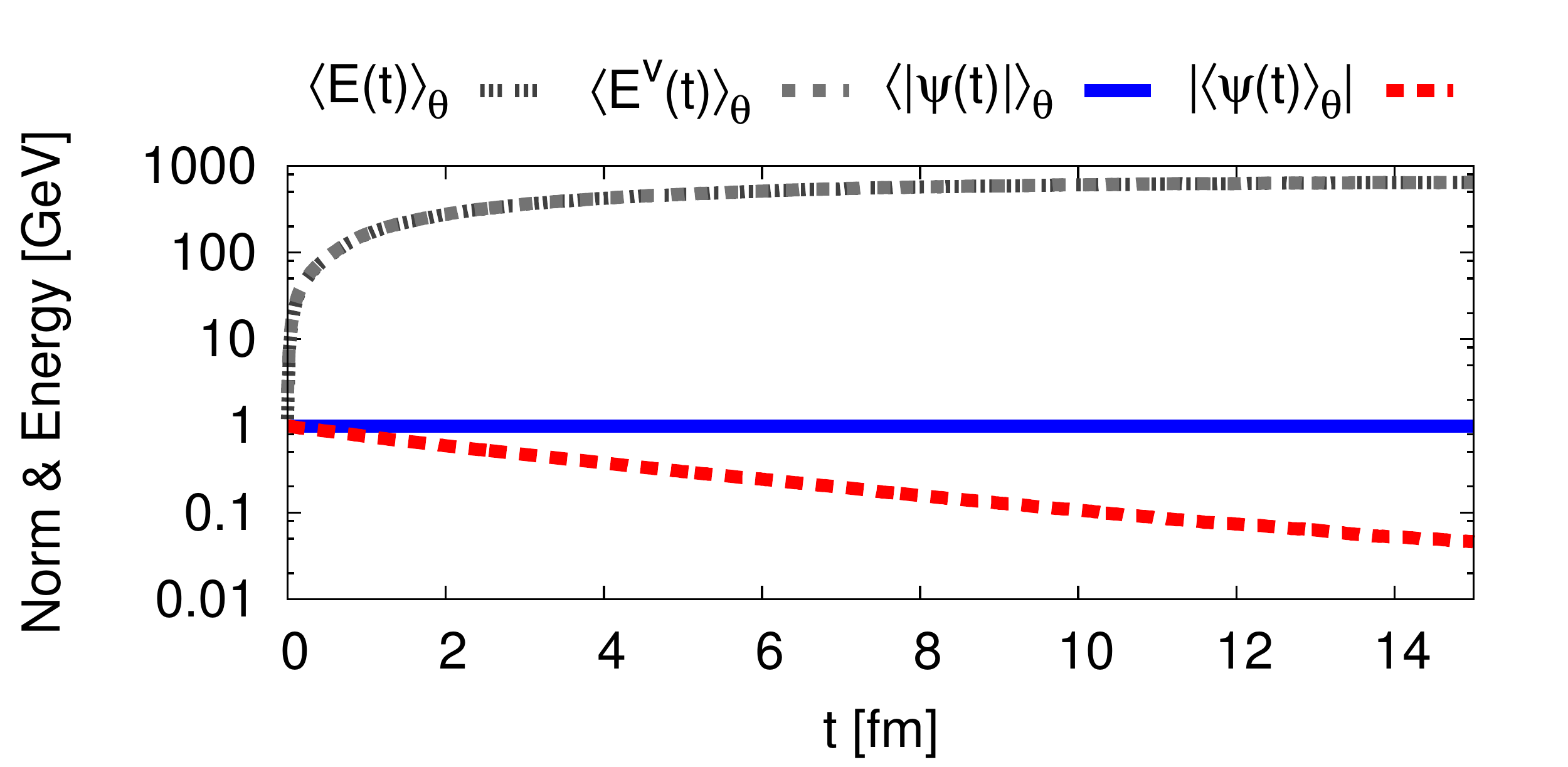}
\caption{ Norm and Energy for the parameter set of model (B). We find unitary evolution, so that the wave function norm $|\psi_{Q\bar{Q}}(t)|=1$ and its energy remains purely real. After averaging, decoherence induces a dampening in $|\langle \psi_{Q\bar{Q}} \rangle_{\theta}(t)|$ as expected from Eq.\eqref{eq:dampingWFavg}. The difference in $E(t)$ and $E^v(t)$ is dominated by the noise contribution to the system energy.}
\label{Eq:Avrgs}
\vspace{-0.4cm}
\label{FigNormNrg}
\end{figure}

To accommodate the full heavy quarkonium wave function Eq.\eqref{eq:schroedinger2} is discretized on a line of $N=512$ points, that covers the physical distance between $x\in[-2.56{\rm fm},2.56{\rm fm}]$. The initial wave function is determined as eigenstate to the Cornell potential $v^{\rm vac}(x)=-\frac{\alpha}{|x|}+\sigma|x|$, with vacuum parameters $\alpha=0.1$ and $\sigma=(0.4 {\rm GeV})^2$. To implement the effects of dynamical quarks, i.e. to account for string breaking, we flatten the linear rise smoothly to a constant value at $r_{\rm sb}=1.5{\rm fm}$.
This choice leads to a total number of $N^{\rm v}_b=4$ bound states. All survival probabilities $P^{\rm v}(t)$  shown in this section are calculated based on the admixture of these four vacuum states contained in the system wave function.  

\begin{figure*}[th!]
\includegraphics[scale=0.35,angle=-90,clip]{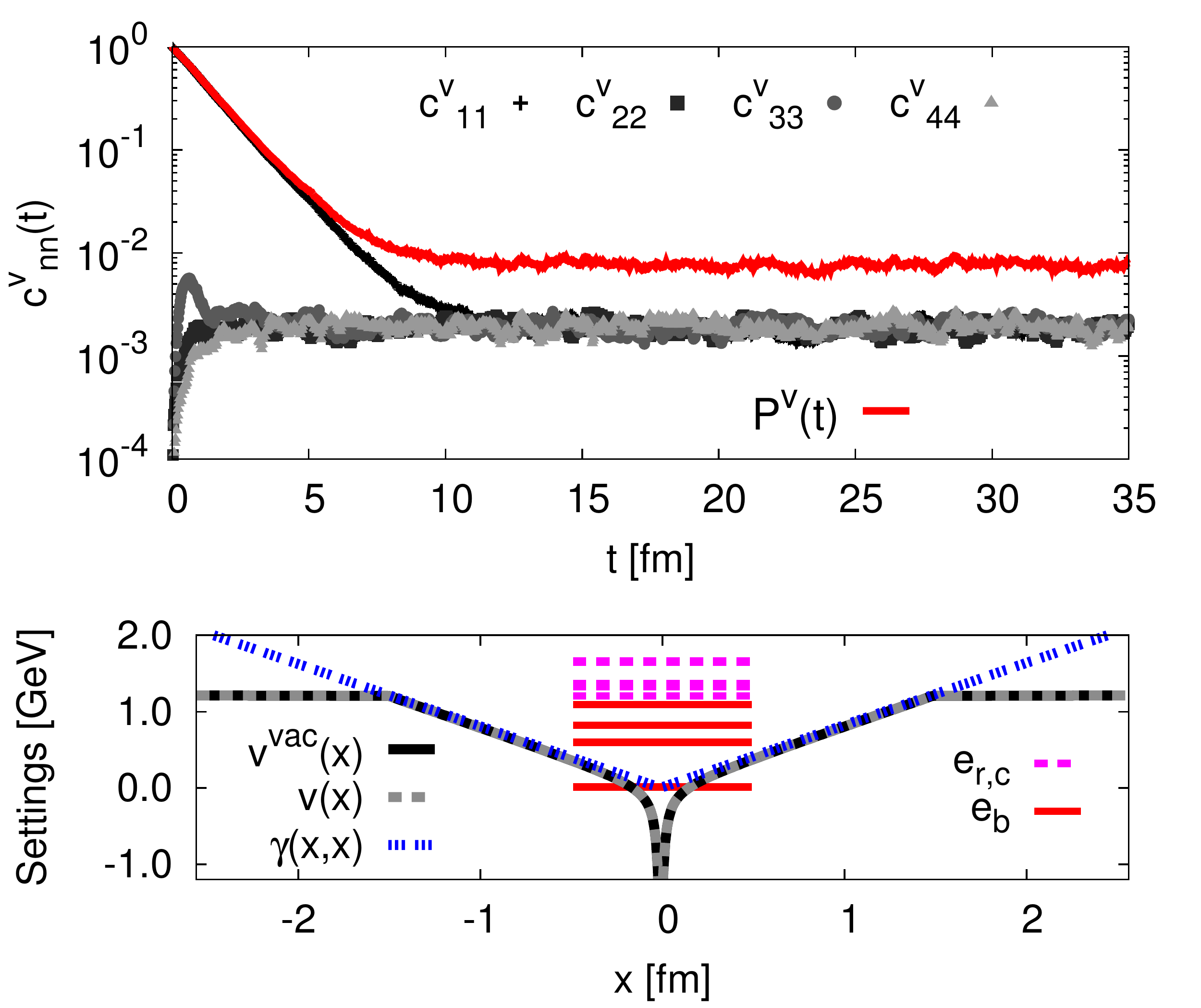}
\includegraphics[scale=0.35,angle=-90,clip]{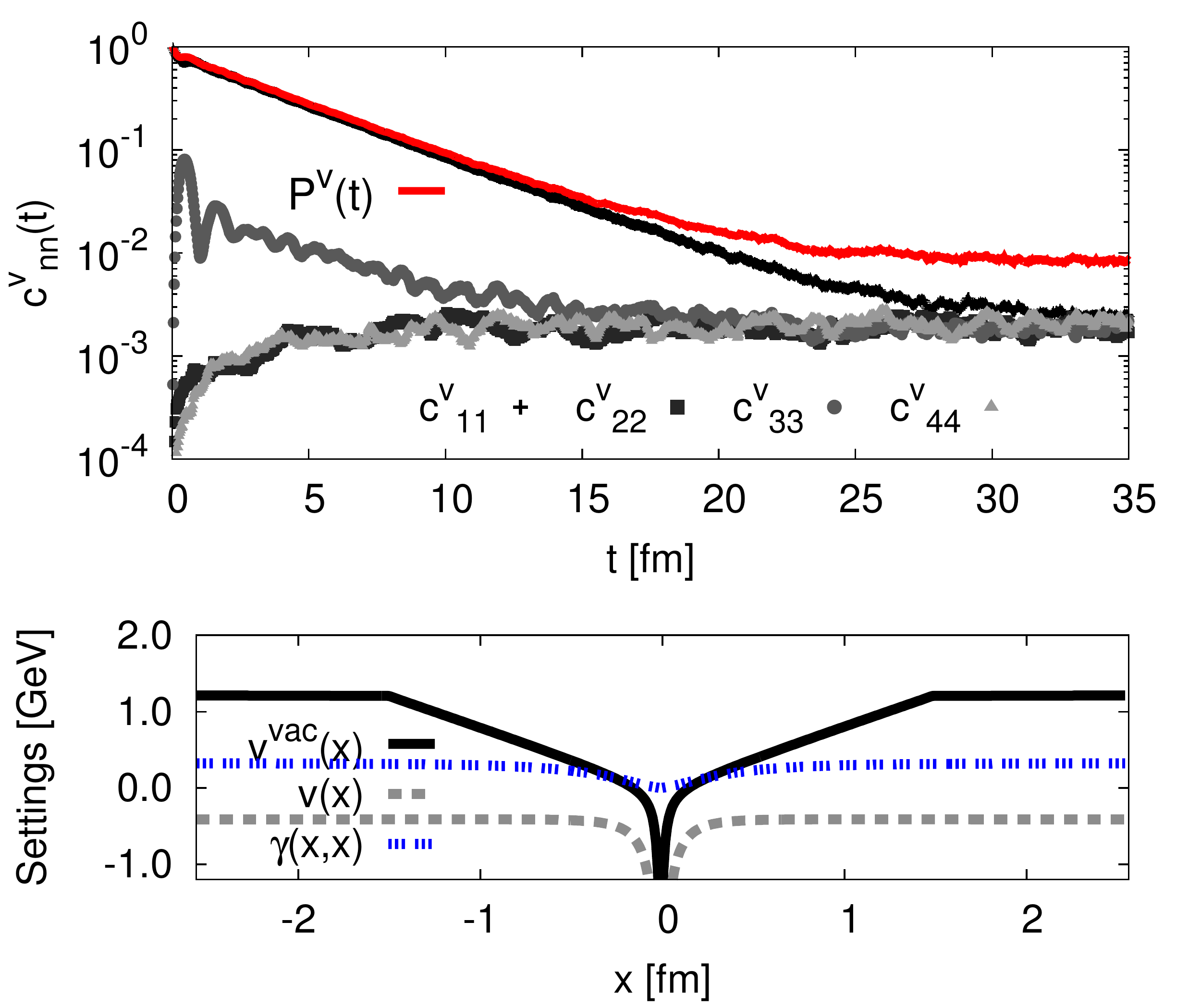}
\caption{One-dimensional model:
(top left) Vacuum admixtures $c^{\rm v}_{nn}(t)$ of heavy quarkonium bound states and their survival probability $P^{\rm v}(t)$ from the stochastic model, based on a lattice QCD inspired parameter set\cite{Rothkopf:2011db}. In the bottom left panel we show the corresponding values of the initial potential $v^{\rm vac}(x)$, the real potential $v(x)$ governing the dynamics and the diagonal noise strength $\gamma(x,x)$. (energy levels of the initial bound states $e_b$ as well as for resonances and continuum solutions $e_{r,c}$ to $v^{\rm vac}(x)$ are given as reference once). Decoherence induces an almost exponential decay of the initial heavy quarkonium ground state at short times before a stable plateau is reached. We find that all available eigenstates $\phi_n$ of $v^{\rm vac}(x)$ are excited and at late times their relative abundances all coincide.
(right) Same quantities with a parameter set adapted from a perturbative (PT) study \cite{Laine:2006ns} where at $T=2.33T_c$ a Debye screened real potential $v(x)={\rm Re}[v^{PT}(x)]$ is accompanied by a small but finite noise term $\gamma(x,x)=2{\rm Im}[v^{PT}(x)]$.}
\label{FigStochDyn1}
\end{figure*}

\begin{figure*}[th!]
\includegraphics[scale=0.35,angle=-90,clip]{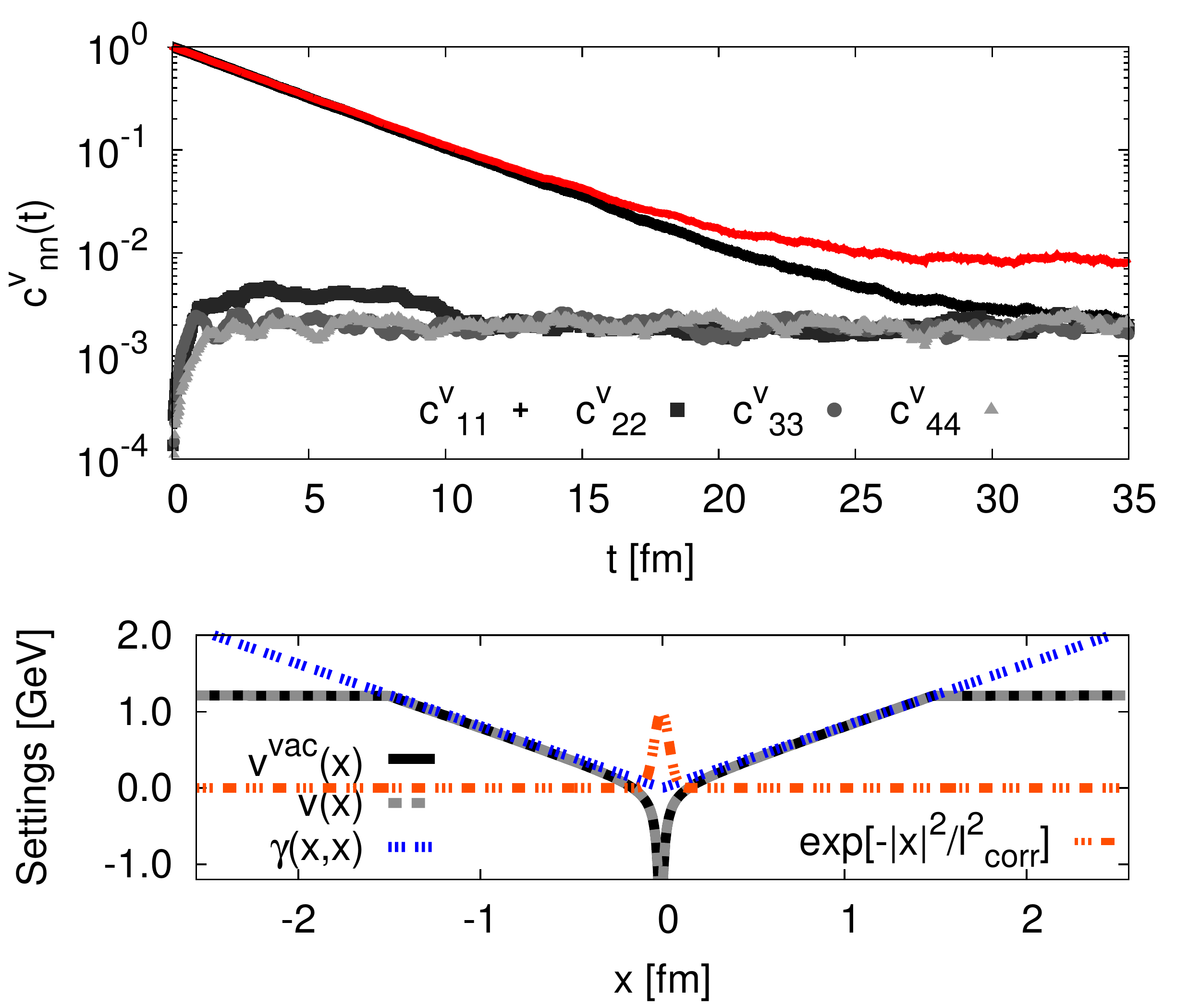}
\includegraphics[scale=0.35,angle=-90,clip]{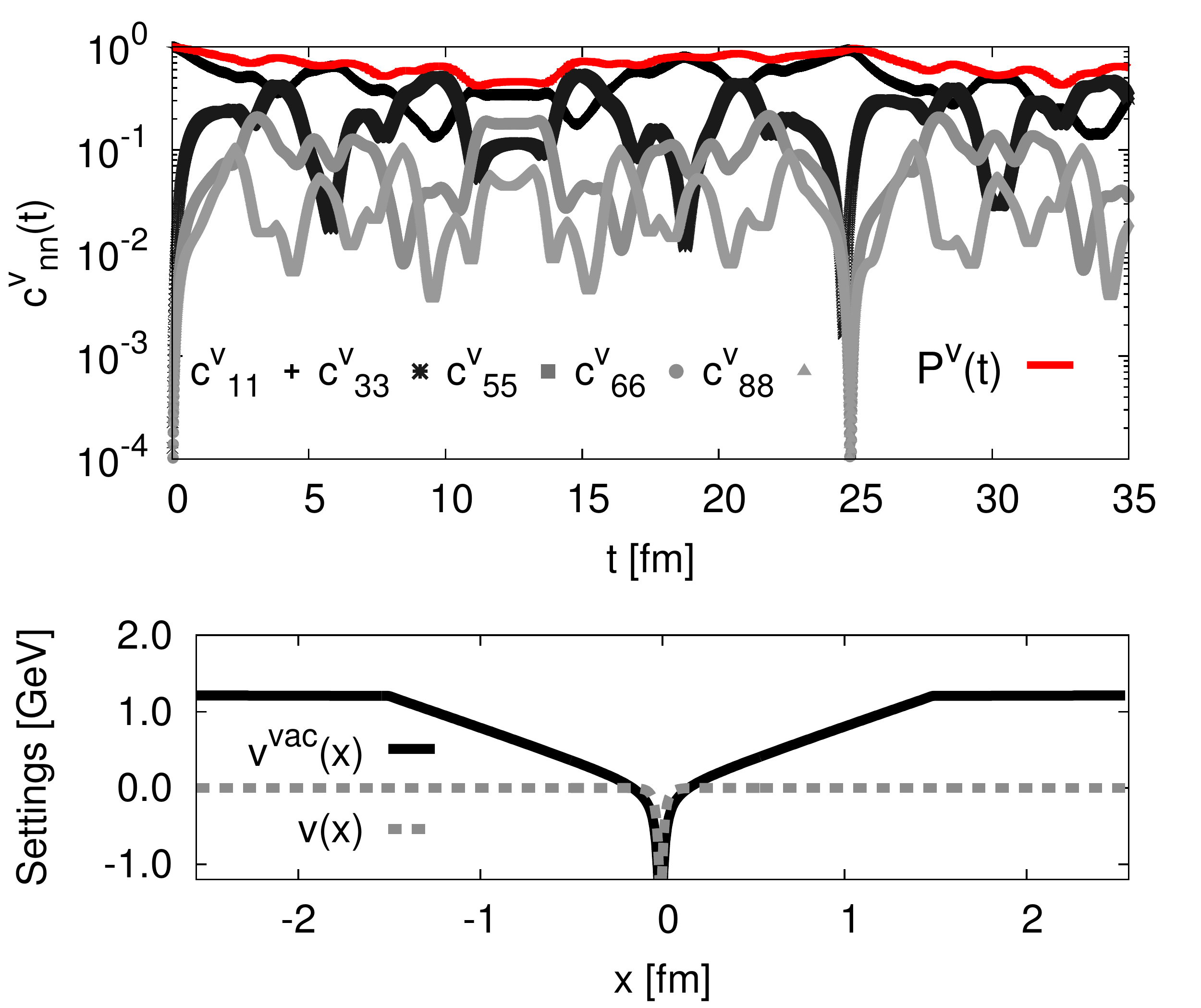}
 \caption{(left) Same $v(x)$ and $\gamma(x,x)$ as in model (A) but with Gaussian damped spatial correlations $\gamma(x,x')\sim\rm{exp}[-|x-x'|^2/l_{corr}^2]$ induced in the noise terms using $l_{\rm corr}=4 {\rm dx}$. 
(right) The classic Debye screening scenario based on a completely real potential $v(x)=-\frac{\alpha e^{-m_D|x|}}{|x|}$ with the Debye mass chosen as $m_D=5{\rm GeV}$. 
}
\label{FigStochDyn2}
\end{figure*}

The stochastic dynamics are implemented using a Crank-Nicholson scheme with time step ${\rm dt}=0.0001{\rm fm}$. There the mass of the constituent quarks is set to $M=1.18 {\rm GeV}$ in order to be larger than the temperature, which we choose to be $T=2.33T_c\sim0.4 {\rm GeV}$. In the first three runs (Fig.\ref{FigStochDyn1} and Fig.\ref{FigStochDyn2} left), off-diagonal noise terms are set to zero, leading to an artificial correlation length $l_{\rm corr}\approx{\rm dx}<l_{\rm th}=15{\rm GeV}^{-1}$. This yields a noise, which is stronger than expected from the physical value of the temperature. 

In the third example (Fig.\ref{FigStochDyn2} left) we approximate the effects of nonvanishing off-diagonal correlations $\gamma( x, x')\propto {\rm exp}(-|x-x'|^2/l^2_{\rm corr})$ by initializing noise in Fourier space
\begin{align}
\nonumber & \langle\theta( p,t)\rangle=0, \quad \langle \theta( p,t)\theta( p',t')\rangle=
\kappa(p)\delta\left( p +  p'\right)\delta_{tt'}/dt \label{Eq:KrnNoise} \\
&\kappa(p)=\int\;dx\;  e^{ipx} \; {\rm exp}\Big(-x^2/l_{\rm corr}^2\Big) 
\end{align}
with a normalized Kernel $\kappa(p)$ at each time step \cite{Namiki:1992wf}. The strength of the diagonal correlations is then set by simple multiplication with $\sqrt{\gamma( x, x)}$ after transforming back to coordinate space.

Before we start a comparison of the different results according to the parameter sets (A)-(D), let us briefly look at generic features of the simulation from the view point of wave function norm 
\begin{align}
 |\psi_{Q\bar{Q}}(t)|^2=\int dx \psi_{Q\bar{Q}}^*(x,t)\psi_{Q\bar{Q}}(x,t)
\end{align}
and energy 
\begin{align}
 \nonumber E(t)&=\int dx \psi_{Q\bar{Q}}^*(x,t) h(x) \psi_{Q\bar{Q}}(x,t)\\
 E^{\rm v}(t)&=\int dx \psi_{Q\bar{Q}}^*(x,t) h^{\rm vac}(x) \psi_{Q\bar{Q}}(x,t)
\end{align}

Both merits and the limitations of the stochastic model of Eq.\eqref{eq:evolution} are already visible in these quantities, which we display in Fig.\ref{Eq:Avrgs} based on the parameters of model (B).

One finds that indeed unitarity is preserved for each individual run of the stochastic ensemble since the wave function norm remains at unity. On the other hand, performing the ensemble average according to Eq.\eqref{eq:dampingWFavg} introduces a dampening of $|\langle \Psi_{Q\bar{Q}}\rangle|$, which is readily observed in the decreasing dashed line in Fig.\ref{Eq:Avrgs}.

Since our stochastic quantum evolution does not include the quantum counterpart of the drag force, momentum of the heavy quarks diffuses in {\it momentum} space without resistance by a friction term.
Consequently one expects that the overall energy of the system over time will increase artificially and indeed one observes a linear rise in the upper two dashed lines in Fig.\ref{FigNormNrg} at intermediate times.
The artificially strong noise pushes the values of the energy to quickly exceed the non-relativistic cutoff $E\sim M\sim 1.18\rm GeV$  and saturates at late times at $E\sim (\pi/{\rm dx})^2/M\sim 3000 {\rm GeV}$ due to the momentum cutoff on the lattice $\pi/{\rm dx}$.
To eliminate the unphysical linear rise in energy and to allow the system to thermalize eventually will require a modification of our approach that takes into account the correct momentum dissipation, i.e. friction as e.g discussed in \cite{Gallis:1991}. Although we do not know how the effect of friction changes the time evolution of the individual heavy quarkonium wave function in detail, we expect these neglected effects to become important once the energy reaches the temperature of the medium (i.e. $E(t)\sim T \sim 0.4 \rm GeV$). Thus the late time behavior observed from the stochastic dynamics presented in this study is to be understood within this limitation. 

Note that especially the simulations of model (A) and (B) are
performed with rather short correlation length ($2\pi/l_{\rm corr} \sim 125 {\rm GeV}$).
Therefore although qualitatively similar behavior should be
observed in simulations with longer correlation
length ($l_{\rm corr}\sim 2\pi/T$), the rapid evolution
presented hereafter must not be understood
as representing a realistic evolution time scale.

{\bf Model (A):} The first run shown on the left of Fig. 1 is based on a parameter set that models thermal effects as fully incorporated into the noise strength. The real potential $v(x)=v^{\rm vac}(x)$ in this case is not modified from its $T=0$ form, while the strength of the fluctuations exhibits a linear rise equal to that of the real part $\gamma(x,x)=\sigma|x|$. 

One finds that the medium via the noise, exponentially decreases the probability to find the initial ground state at early times $t<4 {\rm fm}$. It is the concurrent excitation of higher lying states and their feed-down that at later times $t>7{\rm fm}$ leads to a stabilization of the ground state admixture and thus halting its decline. 

The artificially small correlation length of the noise also appear to render insignificant the energy differences between the individual bound states. Hence already at times $t>2{\rm fm}$ the noise induced admixtures of higher lying states become of equal magnitude. 

Following the time evolution to late times $t>10 \rm fm$ a constant survival probability $P^{\rm v}(t\to\infty)\approx 10^{-2}$ emerges. We however refrain from attaching physical meaning to this feature, since the inclusion of the friction can change the outcome at these times significantly.

{\bf Model (B):} A qualitatively different setup is shown on the right of Fig.1. Its parameters are obtained from an evaluation of the heavy quark potential $v^{\rm PT}(x)$ in resummed perturbation theory \cite{Laine:2006ns}. Here the effects of the medium interaction are present both as modification of the real potential towards a Debye-screened form
\begin{align}
 v(x)={\rm Re}[v^{\rm PT}(x)]\propto -m_D-\frac{e^{-m_D |x|}}{|x|}
\end{align}
with Debye mass $m_D=0.92GeV$, as well as the presence of a noise term
\begin{align}
  \gamma(x,x)=2{\rm Im}[v^{\rm PT}(x)]\propto \int_0^\infty \frac{dz\;z}{(z^2+1)^2}\Big(1-\frac{{\rm sin}[zm_Dx]}{zm_Dx}\Big).
\end{align}

In comparison to model (A), the strength of the noise $\gamma(x,x)$ as well as the real potential are much weaker. We find that in total it takes longer for the ground state to become suppressed. Note that in the absence of noise, the parity even initial ground state is able to mix only with other parity even states under the time evolution of the P-even Hamiltonian $h(x)$. The relatively strong population of the state $\phi_3$ compared to the lighter $\phi_2$ and $\phi_4$ thus tells us that it is such mixing processes that are dominant at times $t<5\rm fm$.

Only after this point in time do the decoherence induced parity odd states contribute to similar strength as $\phi_3$. At late times the asymptotic value for the suppression $P^{\rm v}(t\to\infty)\approx10^{-2}$ in model (B) is very similar to the value observed in model (A), even though its real potential as well as the diagonal noise are very different.

{\bf Model (C):} Let us thus take a look at how spatial correlations in the noise influence the suppression pattern. Purely local noise corresponds to a very high medium temperature if interpreted as a correlation length of the size of the lattice spacing. Thus a realistic description around the deconfinement phase transition will need to allow for larger values of $l_{\rm corr}\sim\frac{2\pi}{T}$. On the left of Fig.\ref{FigStochDyn2} we show the results for a noise with the same local linear rise as in Fig.\ref{FigStochDyn1} left, but with Gaussian damped spatial correlations $\gamma(x,x')\propto \rm{exp}[-|x-x'|^2/l_{corr}^2]$ induced by a convolution with an appropriately normalized Kernel in Fourier space using $l_{\rm corr}=4{\rm dx}$. 

One finds that even though the integrated strength of the noise is the same, i.e. $\frac{1}{N}\sum_p\kappa(p)=1$, its effects are not as localized anymore and thus the suppression of the ground state proceeds slower. The pattern appears to be different also in a qualitative way. We observe e.g. that the excitation of the state $\phi_3$ does not anymore show the overshoot present in all the other scenarios above.

With a noise corresponding to a lower apparent temperature, the difference in mass, i.e. eigenenergy, between the individual bound states becomes more important. Fig.\ref{FigStochDyn2} (left) shows at small times $t<10 \rm fm$ that the amount of population of the states appears to be ordered according to their masses. From the higher lying states, the lightest is consistently excited most strongly. It is the relative abundances of the heavy quarkonia states that are closely related to the off-diagonal components of the noise correlations.

{\bf Model (D):} Taking the route from Model (A) via model (B), we arrive at the Debye screening scenario \cite{Matsui:1986dk}, where all medium effects are modelled as a modification, i.e. a weakening, of the real potential. If real and imaginary part were to each encode part of the thermal effects on the $Q\bar{Q}$ system, the medium effects that were described in model (B) by noise should now be captured by an increased amount of screening present. Hence we choose $m_D= 5 {\rm GeV}$.

As shown on the right of Fig.\ref{FigStochDyn2} one finds that indeed mixing of the eigenstates of $h^{\rm vac}(x)$ occurs and the initial bound state becomes suppressed. $P^{\rm v}(t)$ however does not decrease exponentially not even monotonically.

Note that populating states other than the initial ground state in this case can only be facilitated by mixing through the Hamiltonian. This leads to the observed pattern of the $c_{kk}$'s, where only parity even states are present. (Note that we plot all P-even states up to $\phi_{8}$) 

So far we can summarize our findings: 
\begin{itemize}
 \item The presence of noise leads to spatial decoherence and contributes at short times to an exponential suppression of the initially present heavy quarkonium ground state.
 \item Decoherence also populates the higher lying states. Its effect can differ significantly from the mixing induced by the Hamiltonian $h(x)$ if it is not restricted by selection rules such as e.g. parity. Feed-down from these states leads to a 
 state with respect to $P^{\rm v}(t)$ at late times.
 \item Details of the time evolution, such as suppression speed and the population ratios of heavy quarkonia states already at early times appear to be directly related to the structure of the off-diagonal noise correlations.
 \item The overall suppression at late times in terms of $P^{\rm v}(t)$ is quite insensitive to the details of $v(x)$ and $\gamma(x,x')$. 
This behavior is however an artifact due to neglecting the effect of friction, which becomes important at late times.
\end{itemize} 

\section{Summary and Outlook}\label{sec6}
In this study, we have proposed an open quantum systems approach for the description of $Q\bar Q$ states in a thermal medium using stochastic evolution.
The merits of the stochastic evolution are the possibility 
\begin{itemize}
 \item to give a dynamical description of the non-relativistic heavy $Q\bar{Q}$ evolution in the quark gluon plasma
 \item to give a physical meaning to the existence of an imaginary part in previous studies of the heavy quark potential
 \item to make accessible the concept of spatial decoherence in a potential based description of heavy quarkonium suppression
\end{itemize}
while its applicability is limited to time scales shorter than the heavy quark relaxation time due to
\begin{itemize}
 \item the absence of the friction and the resulting inability to thermalize, expressed in a linearly rising energy $\lim_{t\to\infty}\frac{d}{dt}\langle H \rangle \neq 0$.
\end{itemize}

We have shown how to relate the basic model parameters governing the dynamics, such as $v(r)$ and $\gamma(r,r)$, to the spectral decomposition of the thermal Wilson loop, which in turn can be determined from lattice QCD non-perturbatively.
It is found that the presence of noise leads to spatial decoherence and the consequent dampening of the $Q\bar Q$ wave function, which plays a central role in describing the time evolution of the heavy quark bound states.
One-dimensional model calculations based on Eq.\eqref{eq:schroedinger2} were presented to confirm the viability of our approach.  Their straight forward extension to three dimension thus promises to provide us with new insight on the fate of the heavy quarkonia, such as $J/\Psi$ and $\Upsilon$ in the hot QCD plasma as a function of real-time.

In closing we list several open points of interest as well as possible further improvements:
\begin{enumerate}[(i)]
\item
The color charge is completely ignored in our description.
This is understood as a result of tracing out the color degrees of freedom in the master equation, as we have done in deriving Eq.\eqref{eq:master2} for the density matrix $\hat\rho_{\rm Q\bar Q}(\bm r,\bm r',t)$ in the relative coordinates.
However, it would be more appropriate to give a description of a wave function with color degrees of freedom and derive a more direct connection to QCD.
\item
In addition to quantum decoherence, which we study in this paper and whose classical counterpart is the noise term in the classical Langevin theory, the quantum counterpart of friction is another important ingredient in the description of an open quantum system.
This aspect will become essential in kinetic thermalization and the late time evolution of heavy quark systems as well as resolve the question regarding the monotonous energy increase observed in our numerical simulation.
\item
A first principles definition of $\Gamma(\bm X,\bm X')$ is desirable.
Physical intuition have lead us to relate the thermal wavelength $l_{\rm th}$ and the correlation length of the fluctuation.
Since it gives the length scale of random fluctuation, the entropy change induced by inserting a heavy quarkonium might thus be related to $\Gamma(\bm X,\bm X')$.
It would also be interesting to discuss thermodynamic quantities, such as free energy, in terms of our description. 
\item
Application to relativistic heavy ion collision is interesting and possible in principle, but it would require some phenomenological refinements, e.g. how to combine non-relativistic quantum mechanics with the relativistic hydrodynamic expansion, how to describe hadronization of heavy quarkonium or heavy mesons by using the wave function at the freezeout, and so on.
\end{enumerate}

{\bf Acknowledgment:}
Y.A. is grateful to Masayuki Asakawa, Masakiyo Kitazawa, and Chiho Nonaka for fruitful discussion. 
A.R. acknowledges support through the BMBF project {\em Heavy Quarks as a Bridge between Heavy Ion Collisions and QCD} as well as funding from the Sofja Kovalevskaja program of the Alexander von Humboldt foundation and the EU I3 \textit{Hadron Physics 2}.\vspace{-0.4cm}
% \appendix
\bibliographystyle{apsrev}

\begin{thebibliography}{99}
\bibitem{footnote2}
The spectral function Eq.\eqref{eq:Breit-Wigner} results in divergent integral in Eq.\eqref{eq:continuation}.
The Breit-Wigner form derives from $D^{>}({\bm X},t)\propto \exp\left[-\frac{it}{\hbar}V({\bm X})-\frac{|t|}{2\hbar}\Gamma({\bm{ X,X}})\right]$, which is non-analytic at $t=0$.
The better approximation of $D^{>}({\bm X},t)$ for its short time behavior may affect the spectral function in $|\hbar\omega-V({\bm X})|\gg\Gamma({\bm {X,X}})/2$ and make the integral Eq.\eqref{eq:continuation} convergent.
\end{thebibliography}

\end{document}